\documentclass[conference]{IEEEtran}

\usepackage[usenames,dvipsnames]{color}
\usepackage{subfigure}
\usepackage{graphicx}
\usepackage{amsmath}
\usepackage{color}
\usepackage{multicol}
\usepackage{amssymb}
\usepackage{graphicx}

\usepackage{amsfonts}%
\usepackage{verbatim}%
\usepackage{enumerate}%
\usepackage{psfrag}
\usepackage{epsfig}
\usepackage{epstopdf}
\usepackage{multicol}
\usepackage{setspace}
\usepackage{dsfont}

\usepackage{algorithm}
\usepackage{algorithmic}
\usepackage{cite}
\usepackage{float}
\usepackage{cite}
 
\begin{document}
\title{On Spectrum Sharing Between Energy Harvesting Cognitive Radio Users and Primary Users}
\author{\IEEEauthorblockN{Ahmed El Shafie\IEEEauthorrefmark{1}\IEEEauthorrefmark{4}, Mahmoud Ashour\IEEEauthorrefmark{2}\IEEEauthorrefmark{3}, Tamer Khattab\IEEEauthorrefmark{5} and Amr Mohamed\IEEEauthorrefmark{2}}

\IEEEauthorblockA{\IEEEauthorrefmark{1} Wireless Intelligent Networks Center (WINC), Nile University, Giza, Egypt.\\
\IEEEauthorrefmark{2} Computer Science and Engineering Dept., Qatar University, Doha, Qatar.\\
\IEEEauthorrefmark{5} Electrical Engineering Dept., Qatar University, Doha, Qatar. \\
\IEEEauthorrefmark{4} Electrical Engineering Dept., University of Texas at Dallas, TX, USA. \\
\IEEEauthorrefmark{3} Electrical Engineering Dept., Pennsylvania State University, PA, USA.
}
}
\date{}
\maketitle
\begin{abstract}
This paper investigates the maximum throughput for a rechargeable secondary user (SU) sharing the spectrum with a primary user (PU) plugged to a reliable power supply. The SU maintains a finite energy queue and harvests energy from natural resources and primary radio frequency (RF) transmissions.
We propose a power allocation policy at the PU and analyze its effect on the throughput of both the PU and SU. Furthermore, we study the impact of the bursty arrivals at the PU on the energy harvested by the SU from RF transmissions. Moreover, we investigate the impact of the rate of energy harvesting from natural resources on the SU throughput. We assume fading channels and compute exact closed-form expressions for the energy harvested by the SU under fading. Results reveal that the proposed power allocation policy along with the implemented RF energy harvesting at the SU enhance the throughput of both primary and secondary links.
\end{abstract}
\section{Introduction}


Power supply is considered a crucial aspect affecting the performance of wireless communication systems especially for portable nodes \cite{f1,lu2014dynamic}. Traditionally, portable/mobile wireless nodes are battery-based, i.e., operate with energy supply from a battery. These batteries have limited storage capacity and frequently need to be recharged or replaced \cite{lu2014dynamic}. In the last few years, radio frequency (RF) energy harvesting technology has been developed. Such technology is able to supply energy to wireless nodes through the conversion of RF energy to direct current (DC) energy. The authors of \cite{mikeka2011design} report RF-to-DC conversion efficiencies above 0.4\% at -40 dBm, above 18.2\% at -20dBm and over 50\% at -5 dBm RF signals. There are expectations on the improvement of both the sensitivity of energy harvesting circuits and the conversion efficiency in the near future \cite{lu2014dynamic}. Consequently, the use of RF energy harvesting technology has been gaining increasing world-wide interest. In addition, compared with other forms of energy harvesting (e.g., solar, wind and acoustic noise), RF energy harvesting provides relatively predictable energy supply. The amount of RF harvested energy depends on the wavelength of the harvested RF signal and the channel and distance between an RF energy source and the harvesting device \cite{f3}.

 Several works have considered nodes with energy harvesting capability, e.g., \cite{hoang2009opportunistic,sharma2010optimal,ho2010optimal,yang2010transmission,yang2010optimal,tutuncuoglu2010optimum}. Various energy sources, such as light, vibration or heat are available to be harvested \cite{survey}. In \cite{hoang2009opportunistic}, Hoang \emph{et al.} studied the optimal policies for a cognitive node. The problems of throughput maximization and mean delay minimization for single-node communication were considered in \cite{sharma2010optimal}. Using a finite horizon setup, the authors of \cite{ho2010optimal} addressed the energy allocation for the maximization of the throughput.

Energy harvested from ambient natural energy resources, i.e., wind and solar energy, has got wide attention \cite{pappas,wimob,ourletter,sultan,wcmpaper,gc2013}. The statistics of energy arrivals are assumed to follow certain random processes. For instance, the authors of \cite{pappas,wimob,ourletter,wcmpaper,gc2013} assume Bernoulli energy arrivals. However, the authors of \cite{krikidis2012stability} assume Poisson arrivals. The authors of \cite{le2008efficient,f3,f7,f8,f9} consider energy harvesting from RF transmissions. In particular, it is assumed that nodes can harvest energy from the transmissions of nearby nodes.
Recently, the availability of free RF energy has increased due to the advent of wireless communication and broadcasting systems \cite{le2008efficient}. Radio waves are ubiquitous in our daily lives in
form of signal transmissions from TV, radio, wireless local area networks (LANs) and mobile phones.

Powering a cognitive radio network through RF energy harvesting can be efficient in terms of spectrum usage and energy limits for wireless networking \cite{lu2014dynamic,f7,f8}. In cognitive radio networks with RF energy harvesting capability, secondary users (SUs) harvest energy from RF signals induced from the nearby RF sources, e.g., primary users (PUs), cellular base stations and other ambient RF sources. Such RF signals can be converted into DC electricity. The converted energy can be stored in an energy storage and used to operate the
devices and transmit data. The SUs spend that energy for their own data transmission. Recent literature on RF-powered cognitive radio networks mainly focuses on investigating throughput maximization under
various constraints. The authors of \cite{f9} consider an RF energy harvesting-enabled cognitive radio sensor networks. The total consumed energy should not be greater than the total harvested energy. This condition represents an energy casuality constraint. An optimal mode selection policy is proposed to balance the immediate throughput and harvested RF energy in transmitting and harvesting modes, respectively. The authors in \cite{f8} assume that mobile nodes in a cognitive radio network opportunistically either harvest energy from RF transmissions of nearby devices in a primary network, or transmit data if the secondary nodes are not in the interference range of any PU. The secondary network throughput is maximized via obtaining the optimal transmit power and density of the secondary transmitting terminals under an outage-probability constraint. 

In this work, we consider a mixture between RF energy harvesting and energy harvested from natural resources in the surrounding environment. We assume a simple configuration composed of one energy harvesting SU and a PU. The SU collects energy from nature and converts energy from the PU's RF transmissions. We assume fading channels and propose a power allocation policy at the PU. We investigate the impact of the proposed power allocation policy on the throughput of both primary and secondary links. Results reveal that the throughput of both PU and SU is enhanced under the proposed power allocation and RF energy harvesting techniques.

The rest of this paper is organized as follows. The system model under consideration is described in Section \ref{system_model}. Energy harvesting employed at the SU is studied and analyzed in Section \ref{energy_harvesting}. Section \ref{proposed_policy} presents the proposed transmission policy adopted by the SU along with its throughput analysis. Numerical results are presented in Section \ref{numerical_results}. Finally, concluding remarks are drawn in Section \ref{conclusion}.

\section{System Model}\label{system_model}
\begin{figure}[t]
\begin{center}
\includegraphics[width=1.\columnwidth]{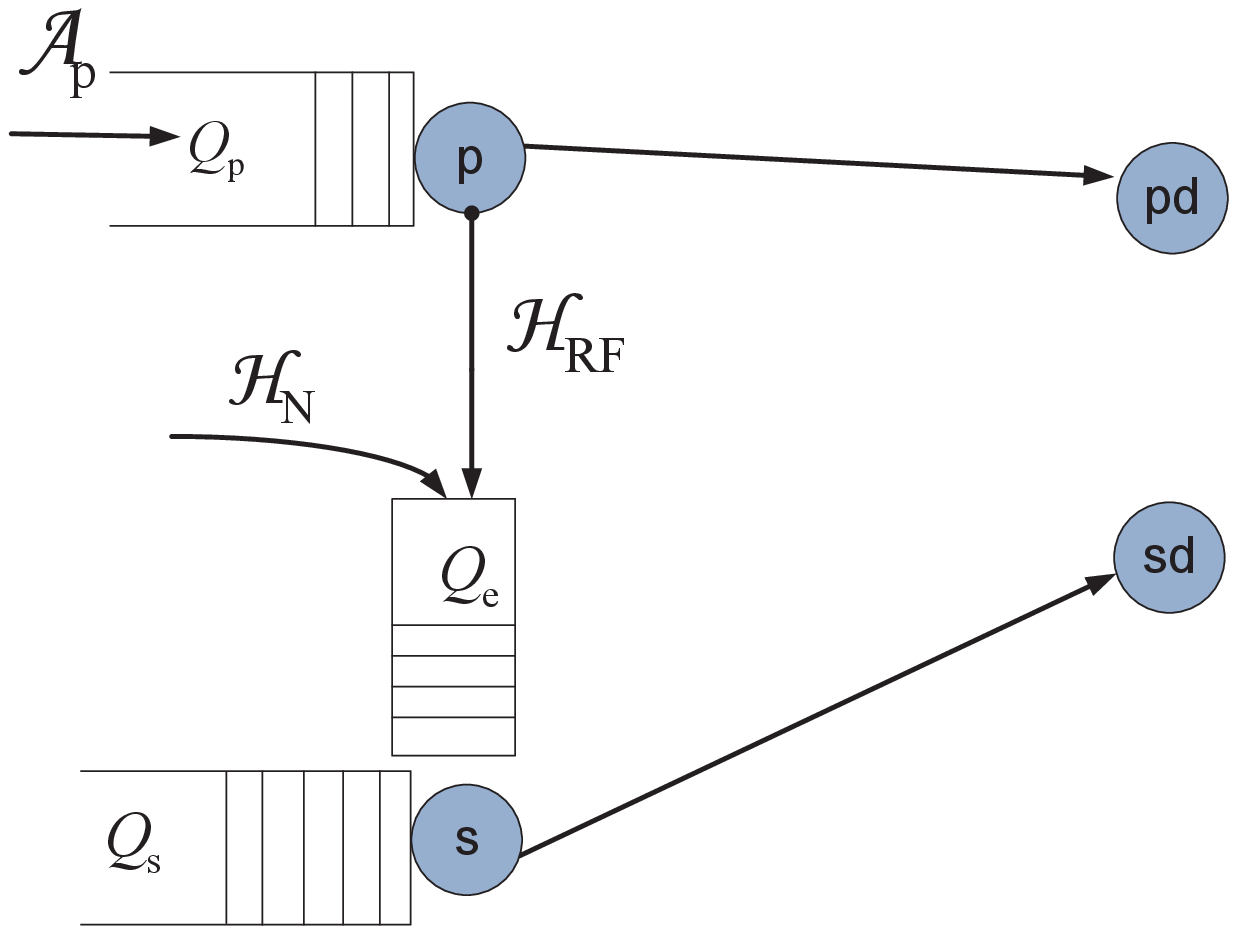}
\caption{Cognitive radio network under consideration. $\mathcal{A}_{\rm p}$ is the number of primary data packet arrivals in a given time slot. $\mathcal{H}_{\rm RF}$ and $\mathcal{H}_{\rm N}$ are the number of energy packets harvested from RF and natural resources, respectively.}\label{Fig1}
\end{center}
\vspace{-5mm}
\end{figure}

Fig. \ref{Fig1} shows the system model under consideration.
We consider a simple cognitive network composed of one PU ($\rm p$) plugged to a reliable power supply and one SU ($\rm s$) equipped with energy harvesting capability.
The PU transmits its packets to a primary destination ($\rm pd$), while the intended destination for SU packet transmissions is $\rm sd$.
Each user has its own infinite-length data queue (buffer) to store the incoming data traffic. We denote the primary and the secondary data queues as $Q_{\rm p}$ and $Q_{\rm s}$, respectively. The SU maintains another queue denoted by $Q_{\rm e}$ which stores fixed-length energy packets. An energy packet is assumed to contain ${\rm e}$ energy units. $Q_{\rm e}$ has a limited capacity of $E_{\max}$ packets. Thus, it contains $E_{\max} {\rm e}$ energy units at maximum. We assume fixed-length data packets of size $\beta$ bits.

Time is slotted and the transmission of a packet takes exactly one time slot of duration $T$ seconds.
The arrivals at the primary queue, $Q_{\rm p}$, are assumed to be independent and identically distributed across time slots following a Bernoulli process with mean $\lambda_{\rm p}\in [0,1]$ packets/slot. We assume that the secondary data queue is saturated with packets. This means that the SU is backlogged with data. The SU harvests energy from the primary RF transmissions and from the environment (nature), e.g., solar, wind, etc. The arrivals to $Q_{\rm e}$ due to environmental energy harvesting are assumed to be distributed according to a Poisson process with rate $\lambda_{\rm e}$ energy packets/slot. For secondary data transmissions, we assume that multiple energy packets may be used for a single data packet transmission. The PU has the priority to transmit if $Q_{\rm p}$ is non-empty, whereas the SU transmits a packet from $Q_{\rm s}$ if $Q_{\rm e}$ maintains at least $\mathcal{G}$ energy packets and the PU is inactive. Whenever it transmits, the PU utilizes the whole time slot duration for its data transmission. However, the SU performs spectrum sensing for a duration of $\tau < T$ seconds to detect the PU's activity. We do not consider erroneous sensing outcomes. For similar assumption of perfect channel sensing, the reader is referred to \cite{krikidis2009protocol,krikidis2010stability,wcmpaper}.

The channel between every transmitter-receiver pair exhibits frequency-flat Rayleigh block fading, i.e., the channel coefficient remains constant for one time slot and changes independently from a slot to another. We denote the gain of the channel between transmitter ${\rm i}$ and receiver ${\rm \ell}$ at the $t$th time slot by $h_{\rm i\ell}(t)$\footnote{We refer by the channel gain to the absolute squared value of the channel fading coefficient.}. According to the Rayleigh fading assumption, $h_{\rm i\ell}(t)$ is exponentially distributed with mean $\rm \sigma_{i\ell}$. All links are statistically independent.

The PU adopts a power allocation scheme in which it selects its transmission power level each time slot based on the channel realization between its transmitter and the destination. For the link $\rm p \rightarrow pd$, we assume perfect channel state information at the transmitter side (CSIT).
The transmit power is assumed to be continuous over the set $P=[0,P_\mathcal{M}]$, where $P_\mathcal{M}$ is the maximum transmit power constraint per time slot and is dependent on the application and the used transceiver. If the primary direct link is in a deep fade such that the maximum power, $P_\mathcal{M}$, cannot overcome the link outage, the PU remains idle during the whole time slot to avoid wasting its energy without adding further contribution to its throughput. An outage occurs on the link $\rm p \rightarrow pd$ when the transmission rate exceeds the link capacity. The probability that an outage occurs on that link in the $t$th time slot is given by
\begin{equation}\label{outage_p}
\mathbb{P}_{\rm ppd}(t)=\Pr \left \{\mathcal{R}_{\rm p}> \log_2\left(1+\frac{P_{\rm p}(t) h_{\rm ppd}(t)}{\mathcal{N}_\circ W}\right)\right \}
\end{equation}
where $\Pr\{\cdot\}$ denotes the probability of the event $\{\cdot\}$, $\mathcal{R}_{\rm p}=\beta/T/W$ bits/second/Hz is the targeted primary spectral efficiency, $P_{\rm p}(t)$ is the average transmit power in Watts used by the PU in time slot $t$, ${\mathcal{N}_\circ}$ is the additive white Gaussian noise (AWGN) power spectral density in Watts/Hz and $W$ is the channel bandwidth in Hz. It is worth noting from (\ref{outage_p}) that the minimum power required to guarantee a successful primary transmission in time slot $t$, i.e., no outage occurs on the link $\rm p \rightarrow pd$, is given by
\begin{equation}\label{PU_power}
P^*_{\rm p}(t)=\frac{\mathcal{N}_\circ W (2^{\mathcal{R}_{\rm p}}-1)}{h_{\rm ppd}(t)}.
\end{equation}
Hereafter, we omit the temporal index $t$ for simplicity. Nevertheless, it is implicitly understood that power allocation at the PU is performed on a slot-by-slot basis.
The PU sets its transmission power level according to (\ref{PU_power}) and $P_{\mathcal{M}}$. If $P_{\rm p}^*>P_{\mathcal{M}}$, the PU remains silent. It is worth noting that as the instantaneous channel gain, $h_{\rm ppd}$, decreases, the power level used by the PU increases. This potentially increases the amount of RF energy harvested by the SU. However, as $h_{\rm ppd}$ decreases, this also increases the probability of the PU being idle which in turn reduces its throughput and might result in increasing the availability of time slots in which the SU can transmit. Therefore, this point arises a fundamental tradeoff between the throughput of both the PU and SU which is studied in details in the context of the paper. Interestingly, results reveal that the throughput of both users could be enhanced simultaneously.

\section{Energy harvesting}\label{energy_harvesting}
In this section, we study and analyze the arrivals of energy packets to $Q_{\rm e}$ as a result of harvesting energy from primary RF transmissions as well as from natural resources. We lay down the basis of the probabilistic energy arrivals model based on which the analysis in the rest of the paper is performed.

\subsection{RF energy harvesting}
The one and only source of RF energy harvesting considered in this paper is the transmission of PU's packets. Therefore, to analyze the impact of these transmissions on the amount of harvested energy by the SU, we start with studying the behavior of the PU. The SU accesses the channel when the PU is inactive. This occurs in two cases. 1) $Q_{\rm p}$ is empty; or 2) $Q_{\rm p}$ is non-empty but $P_{\rm p}^{*}>P_{\mathcal{M}}$. Thus, the probability of the PU being inactive is given by
\begin{equation}
\begin{split}
\Pi\!&=\!\Pr\{Q_{\rm p}=0\}\!+\!\Pr\{Q_{\rm p}\!\ne\!0\} \Pr\{P_{\rm p}^*>P_{\mathcal{M}}\}\!
\\& =\!1\!-\!\Pr\{Q_{\rm p}\ne0\} (1-\Pr\{P_{\rm p}^*\!>\!P_{\mathcal{M}}\}).
\end{split}
\end{equation}
We first compute $\Pr\{P_{\rm p}^*>P_{\mathcal{M}}\}$. It is given by
\begin{equation}
\Pr\{P_{\rm p}^*>P_{\mathcal{M}}\}= \Pr \left \{\frac{\mathcal{N}_\circ W(2^{\mathcal{R}_{\rm p}}-1)}{h_{\rm ppd}}>P_{\mathcal{M}} \right \}.
\end{equation}
Since $h_{\rm ppd}$ is exponentially distributed, we have
\begin{equation}
\Pr\{P_{\rm p}^*>P_{\mathcal{M}}\}= 1-\exp\Big(-\frac{\mathcal{N}_\circ W(2^{\mathcal{R}_{\rm p}}-1)}{P_{\mathcal{M}}\sigma_{\rm ppd}}\Big).
\end{equation}

The mean service rate of the primary queue, $\mu_{p}$, is equal to the probability that the link $\rm p \rightarrow pd$ is not in outage. Therefore, $\mu_{p}$ is given by
\begin{equation}\label{mu_p}
\begin{split}
&\mu_{\rm p}=\Pr\{P_{\rm p}^* \leq P_{\mathcal{M}}\}=\exp\Big(-\frac{\mathcal{N}_\circ W (2^{\mathcal{R}_{\rm p}}-1)}{P_{\mathcal{M}}\sigma_{\rm ppd}}\Big).
\end{split}
\end{equation}
If $Q_{\rm p}$ is stable, i.e., $\lambda_{\rm p}<\mu_{\rm p}$, the probability of the queue being empty is given by
 \begin{equation}
\Pr\{Q_{\rm p}=0\}=1-\frac{\lambda_{\rm p}}{\mu_{\rm p}}.
\end{equation}
However, if $Q_{\rm p}$ is unstable, i.e., $\lambda_{\rm p}\ge\mu_{\rm p}$, the queue is always saturated with data packets and hence, the probability of the queue being empty is $0$.
Combining both cases, the probability of the primary queue being empty is given by
\begin{equation}
\Pr\{Q_{\rm p}=0\}=1-\min \left \{ \frac{\lambda_{\rm p}}{\mu_{\rm p}},1 \right \}.
\end{equation}
Thus, the probability of the PU being inactive is given by
\begin{equation}\label{Pi}
\Pi=1-\min \left \{ \frac{\lambda_{\rm p}}{\mu_{\rm p}},1 \right \} \mu_{\rm p}
\end{equation}
where $\mu_{\rm p}$ is given by (\ref{mu_p}).
Consequently, the throughput of the PU is given by
\begin{equation}\label{PU_thrpt}
\overline{\Pi}=\min \left \{ \frac{\lambda_{\rm p}}{\mu_{\rm p}},1 \right \} \mu_{\rm p}.
\end{equation}
where the notation $\overline{\mathcal{X}}=1-\mathcal{X}$ is used throughout the paper.

The received power at the SU due to the primary transmission, when the PU is {\it active}, is given by
\begin{equation}\label{rxed_p_su}
P_{\rm p}^* h_{\rm ps} = \frac{\mathcal{N}_\circ W (2^{\mathcal{R}_{\rm p}}-1)}{h_{\rm ppd}} h_{\rm ps}
\end{equation}
with $P_{\rm p}^*\le P_{\mathcal{M}}$. Received signals from primary transmissions at the SU is converted into energy packets with an RF-to-DC conversion efficiency $\eta \leq 1$. From (\ref{rxed_p_su}), it is obvious that the amount of energy harvested by the SU depends on the quality of the links between the PU and SU and the PU and its destination, i.e., $\rm p \rightarrow s$ and $\rm p \rightarrow pd$. Another interesting observation is the dependence of the harvested energy on the burstiness of the primary source. Specifically, more data packets at the PU increases the possibility of transmission and hence, increases the possibility of harvesting energy at the SU. However, this comes at the expense of lowering the possibility of finding available time slots in which the SU can transmit.

Since one energy packet contains ${\rm e}$ energy units, the number of energy packets harvested by the SU when the PU is {\it active} in a time slot is given by
\begin{equation}
\mathcal{H}_{\rm RF}=\Big\lfloor \frac{\eta  \frac{\mathcal{N}_\circ W (2^{\mathcal{R}_{\rm p}}-1)}{h_{\rm ppd}} h_{\rm ps}T}{{\rm e}}\Big\rfloor
\end{equation}
where $\lfloor\cdot\rfloor$ is the largest integer not greater than the argument.
Therefore, when the PU transmits in a given time slot, the probability that the SU harvests $n$ energy packets is given by
\begin{equation}\label{P(A_e=n)}
\Pr \lbrace \mathcal{H}_{\rm RF}\!=\!n \rbrace\!=\! \Pr \! \left \lbrace \!\! n \leq \frac{\eta \frac{\mathcal{N}_\circ W(2^{\mathcal{R}_{\rm p}}-1)}{h_{\rm ppd}} h_{\rm ps}T}{{\rm e}}
< n+1 \!\! \right \rbrace
\end{equation}
with $h_{\rm ppd} \geq a=\frac{\mathcal{N}_\circ W(2^{\mathcal{R}_{\rm p}}-1)}{P_{\mathcal{M}}}$. This condition on $h_{\rm ppd}$ originates from the power allocation policy employed at the PU. From (\ref{P(A_e=n)}), $\Pr \lbrace \mathcal{H}_{\rm RF}\!=\!n \rbrace$ can be written as
\begin{equation}\label{Z}
\Pr \left \{ \frac{n\rm e}{\eta \mathcal{N}_\circ W(2^{\mathcal{R}_{\rm p}}-1)T} \leq
\frac{h_{\rm ps}}{h_{\rm ppd}}
< \frac{(n+1){\rm e}}{\eta \mathcal{N}_\circ W(2^{\mathcal{R}_{\rm p}}-1)T}\right \}.
\end{equation}
Let $\alpha=\frac{{\rm e}}{\eta {\mathcal{N}_\circ W(2^{\mathcal{R}_{\rm p}}-1)} T}$. We solve (\ref{Z}) to obtain a closed-form expression for the probability of harvesting $n$ packets through primary RF transmissions. We get
\begin{equation}
\Pr \lbrace \mathcal{H}_{\rm RF}\!=\!n \rbrace = F((n+1)\alpha)-F(n\alpha)
\end{equation}
where the function $F(\cdot)$ is derived in Appendix A.

\subsection{Energy harvesting from natural resources}
The energy packets harvested from nature are assumed to follow a Poisson process with rate $\lambda_{\rm e}$ \cite{krikidis2012stability}. Thus, the probability of harvesting $k$ energy packets in a given time slot is given by
\begin{equation}
\Pr\{\mathcal{H}_{\rm N}=k\}=\frac{(\lambda_{\rm e} T)^k \exp(-\lambda_{\rm e} T)}{k!}.
\end{equation}

Note that the number of harvested packets from any of the sources cannot be negative; hence, $P^{\prime}_{n}=P^{\prime \prime}_{n}=0$ for all $n\in\{-\infty,\dots,-2,-1\}$.
\subsection{Combining nature and RF energy harvesting}
When the PU is inactive, the energy packet arrivals at $Q_{\rm e}$ comes from natural resources only. Thus, the probability of harvesting $n$ packets of energy from nature is given by
\begin{equation}
P^{\prime}_{n}= \Pr \{\mathcal{H}_{\rm N}=n\}.
\label{dfgg2}
\end{equation}
However, when the PU is active, the probability of harvesting $n$ packets of energy from both the PU's transmission and natural resources in an arbitrary time slot is given by
\begin{equation}
P^{\prime \prime}_{n}=\sum_{k=0}^{\infty}\Bigg(\begin{array}{cc}
                     n \\
                    k
                  \end{array}\Bigg)
\Pr \{\mathcal{H}_{\rm N}=k\} \Pr \{\mathcal{H}_{\rm RF}=n-k\}.
\label{dfgg}
\end{equation}
where $\Big(\begin{array}{cc}
                     n \\
                    k
                  \end{array}\Big)$ denotes $n$ choose $k$. Note that the joint probabilities in (\ref{dfgg}) are the multiplication of the marginal probabilities due to the independence of the two energy sources, i.e., RF and natural energy sources.

The energy queue evolves as follows:
   \begin{equation}
Q^{t+1}_{\rm e}=(Q^{t}_{\rm e}- \mathcal{D}^t_{\rm e})^+  +\mathcal{H}^{t}_{\rm RF}+\mathcal{H}^{t}_{\rm N}
\end{equation}
where $(\cdot)^+$ returns the maximum between the argument and $0$, $\mathcal{D}^t_{\rm e}$ is the number of departures from the energy queue in time slot $t$, $\mathcal{H}^t_{\rm RF}$ is the number of harvested energy packets from the primary transmission in time slot $t$, and $\mathcal{H}^t_{\rm N}$ is the number of energy harvested from nature in time slot $t$.

\section{Proposed Transmission Policy and SU throughput analysis}\label{proposed_policy}
Whenever the SU transmits a data packet, it uses $\mathcal{G}~\in~\{1,2,3,\dots,E_{\max}\}$ energy packets. This means that once that number of packets is available at the energy queue and at the same time the PU is inactive, they are dissipated. Clearly, as the number of energy packets invested in a secondary data packet transmission increases, i.e., as $\mathcal{G}$ increases, the probability that $\rm sd$ successfully decodes that packet increases. However, increasing $\mathcal{G}$ depletes $Q_{\rm e}$ faster. Therefore, we need to choose the optimal value of $\mathcal{G}$ that maximizes the SU throughput.

Let $\chi_m$ denote the probability of the energy queue being in state $m\in\{0,1,\dots,E_{\max}\}$.
Therefore, it can be easily shown that the mean service rate of the energy queue is given by
\begin{equation}
\begin{split}
\mu_{\rm e}&= \mathcal{G} \Pi  \sum_{j=\mathcal{G}}^{E_{\max}}  \chi_j
\end{split}
\end{equation}
where $\Pi$ is given by (\ref{Pi}).
The SU transmits with fixed energy $\mathcal{G} {\rm e}$ energy units. Each time slot, it senses the channel for $\tau$ seconds and transmits only if the channel is sensed to be free. Thus, its transmission spectral efficiency is $\mathcal{R}_{\rm s}=b/T_{\rm s}/W$ bits/second/Hz, where $T_{\rm s}=T-\tau$. The probability of success of a secondary data packet transmission, i.e., the link $\rm s \rightarrow sd$ is not in outage, is given by
\begin{equation}\label{P_ssd}
\overline{\mathbb{P}_{\rm ssd,\mathcal{G}}}= \exp\Big(-\mathcal{N}_\circ W(T-\tau)\frac{(2^{\mathcal{R}_{\rm s}}-1)}{\mathcal{G} {\rm e}\sigma_{\rm ssd}}\Big).
\end{equation}
As indicated earlier, $\overline{\mathbb{P}_{\rm ssd,\mathcal{G}}}$ is monotonically increasing with the transmit energy, $\mathcal{G} {\rm e}$.

A packet from the secondary data queue is served when the PU is inactive, the secondary link is not in outage and the secondary energy queue maintains $\mathcal{G}$ energy packets. The throughput of the SU is then given by
\begin{equation}\label{SU_thrpt}
\begin{split}
\mu_{\rm s}&=\Pi \overline{\mathbb{P}_{{\rm ssd},\mathcal{G}}} \sum_{j=\mathcal{G}}^{E_{\max}}  \chi_j.
\end{split}
\end{equation}
The maximum SU throughput is obtained via solving the following optimization problem:
\begin{equation}
\begin{split}
& \underset{\mathcal{G}\in \{1,2,\dots,E_{\max}\}}{\max.} \,\,\,\,\,\,\ \mu_{\rm s}=\Pi    \overline{\mathbb{P}_{{\rm ssd},\mathcal{G}}} \sum_{j=\mathcal{G}}^{E_{\max}}  \chi_j.
\end{split}
\end{equation}
$\Pi$ and $\overline{\mathbb{P}_{{\rm ssd},\mathcal{G}}}$ have already been determined by (\ref{Pi}) and (\ref{P_ssd}), respectively. Thus, it now remains to compute the steady state distribution of the energy queue, i.e., $\{\chi_j\}_{j=\mathcal{G}}^{\infty}$, to completely characterize the SU throughput for a given value of $\mathcal{G}$, given by (\ref{SU_thrpt}). Towards this objective, we model the evolution of the energy queue with a Markov chain. The probability of transition from state $j$ to state $k$, denoted by $P_{j \rightarrow k}$, is
\begin{itemize}
\item For $k < E_{\max}$,
\begin{enumerate}
\item at $j<\mathcal{G}$,
\begin{enumerate}
\item if $k<j$,
\begin{equation}
P_{j \rightarrow k}=0
\end{equation}
\item if $k \geq j$,
\begin{equation}
P_{j \rightarrow k}=\Pi P^{\prime}_{k-j} +
\overline{\Pi}P^{\prime \prime}_{k-j}
\end{equation}
\end{enumerate}
\item at $j \geq \mathcal{G}$,
\begin{enumerate}
\item if $k<j-\mathcal{G}$,
\begin{equation}
P_{j \rightarrow k}=0
\end{equation}
\item if $k \geq j-\mathcal{G}$,
\begin{equation}
P_{j \rightarrow k}=\Pi P^{\prime}_{k-(j-\mathcal{G})} +
\overline{\Pi}P^{\prime \prime}_{k-j}.
\end{equation}
\end{enumerate}
\end{enumerate}
\item For $k=E_{\max}$,
\begin{enumerate}
\item at $j<\mathcal{G}$,
\begin{align}
P_{j \rightarrow k}&=\Pi \left[\displaystyle \sum_{n=k-j}^{\infty} P_{n}^{\prime}\right] +
\overline{\Pi} \left[\displaystyle \sum_{n=k-j}^{\infty} P_{n}^{\prime \prime}\right] \notag \\
&= \Pi \left[ 1-\displaystyle \sum_{n=0}^{k-j-1} P_{n}^{\prime} \right] +
\overline{\Pi} \left[1- \displaystyle \sum_{n=0}^{k-j-1} P_{n}^{\prime \prime} \right]
\end{align}
\item at $j \geq \mathcal{G}$,
\begin{align}
P_{j \rightarrow k}&=\Pi \left[\displaystyle \sum_{n=k-(j-\mathcal{G})}^{\infty} P_{n}^{\prime}\right] +
\overline{\Pi} \left[ \displaystyle \sum_{n=k-j}^{\infty} P_{n}^{\prime \prime} \right] \notag \\
&= \Pi \left[ 1-\displaystyle \sum_{n=0}^{k-(j-\mathcal{G})-1} P_{n}^{\prime} \right] +
\overline{\Pi} \left[1- \displaystyle \sum_{n=0}^{k-j-1} P_{n}^{\prime \prime} \right].
\end{align}
\end{enumerate}
\end{itemize}
\begin{figure}[t]
\begin{center}
\includegraphics[width=1\columnwidth , height=0.7\columnwidth]{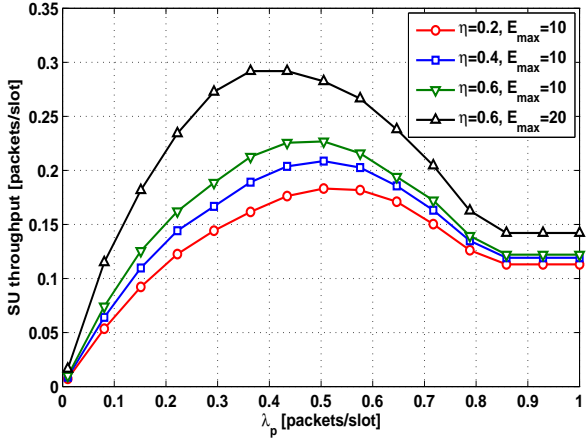}
\caption{SU throughput versus $\lambda_{p}$ for different values of $\eta$ and $E_{\max}$.} \label{Fig2}
\end{center}
\vspace{-5mm}
\end{figure}

The $(E_{\max}+1) \times (E_{\max}+1)$ transition probability matrix, denoted by $\Omega$, is constructed according to the above description. Then, the $1 \times E_{\max}$ steady state distribution vector $\boldsymbol{\chi}=[\chi_0,\chi_1,\dots,\chi_{E_{\max}}]$ is obtained through solving
\begin{equation}
\boldsymbol{\chi}=\boldsymbol{\chi} \Omega.
\end{equation}

\section{Numerical Results}\label{numerical_results}
In this section, the system performance is evaluated in terms of the SU throughput under the proposed transmission policy. Specifically, we investigate the effect of the following factors on the achievable throughput for the secondary link $\rm s \rightarrow sd$;
\begin{enumerate}
\item We study the impact of the bursty arrivals at the PU's queue on the SU throughput.
\item We analyze the dependence of both primary and secondary throughput on the quality of the direct link between the PU and its destination, i.e., we focus on the role of the power allocation policy employed at the PU.
\item We show how each energy harvesting technique employed at the SU influences its throughput, i.e., we investigate the effect of harvesting energy from natural resources and PU's RF transmissions individually. Then, we combine both energy harvesting techniques to enhance the secondary throughput.
\item The roles of the maximum capacity of the energy queue, $E_{\max}$, energy packets' arrival rate at $Q_{\rm e}$, $\lambda_{\rm e}$, and RF-to-DC conversion efficiency, $\eta$, are highlighted.
\end{enumerate}

We show our results at a fixed packet-length of $\beta=10^3$ bits. Time slot duration $T$ is set to $1$ second, while the sensing duration $\tau=0.1$ second. We note that $10 \%$ of the time slot is used by the SU to perform sensing which to some extent justifies the assumption of perfect sensing.
Channel bandwidth $W=1$ KHz. The noise power spectral density $\mathcal{N}_{\circ}=10^{-6}$ Watts/Hz. We assume that an energy packet contains $10^{-3}$ joules, i.e., ${\rm e}=10^{-3}$ joules. The maximum power constraint imposed at the PU $P_\mathcal{M}=10$ dBm unless otherwise stated. The variances of both links $\rm p \rightarrow s$ and $\rm s \rightarrow sd$ are set to unity, i.e., $\rm \sigma_{ps}=\sigma_{ssd}=1$. However, we consider $\rm \sigma_{ppd}=0.5$.

\begin{figure}[t]
\begin{center}
\includegraphics[width=1\columnwidth , height=0.7\columnwidth]{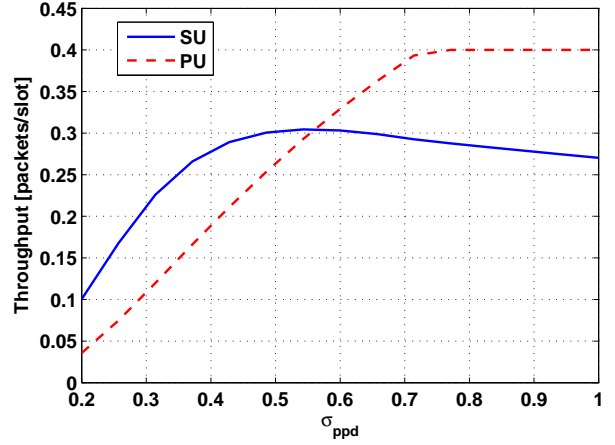}
\caption{Primary and secondary throughput versus $\sigma_{\rm ppd}$.} \label{Fig3}
\end{center}
\vspace{-5mm}
\end{figure}

In Figs. \ref{Fig2} and \ref{Fig3}, we demonstrate the effect of RF energy harvesting on the system performance. We consider a scenario in which the SU harvests energy from the PU's RF transmissions only, i.e., $\lambda_{\rm e}=0$. We study the effect of the packet arrival rate at $Q_{\rm p}$, $\lambda_{\rm p}$, on the maximum achievable SU throughput. Towards this objective, we plot in Fig. \ref{Fig2} the maximum SU throughput versus $\lambda_{\rm p}$. Interestingly, we observe that the SU throughput is initially enhanced as $\lambda_{\rm p}$ increases. However, this behavior is reversed as $\lambda_{\rm p}$ continues to increase until the secondary throughput saturates as $\lambda_{\rm p} \in [\mu_{\rm p},1]$. What happens beyond the scene goes as follows. Initially, as $\lambda_{\rm p}$ increases, the PU transmits more frequently and hence, the amount of energy harvested by the SU due to primary RF transmissions increases. Thus, the SU possesses enough energy for its prospective transmissions which enhances its throughput. However, as $\lambda_{\rm p}$ continues to increase, the SU suffers from the lack of available time slots in which it can transmit, i.e., slots in which the PU is inactive. This becomes the dominant factor reducing the throughput of the SU even though it has enough amount of energy. As $\lambda_{\rm p}$ exceeds $\mu_{\rm p}$, the PU's activity no longer depends on $\lambda_{\rm p}$ since its queue becomes backlogged. From (\ref{PU_thrpt}), we know that $\overline{\Pi}=\mu_{\rm p}$ which is a constant number independent of $\lambda_{\rm p}$. Therefore, the effect of $\lambda_{\rm p}$ on the SU throughput in the interval $\lambda_{\rm p} \in
[\mu_{\rm p},1]$ is neutralized and the secondary throughput saturates as shown in Fig. \ref{Fig2}.

Furthermore, we depict the impact of RF-to-DC conversion efficiency on the SU throughput via plotting the results for different values of $\eta$ in Fig. \ref{Fig2}. Obviously, the SU throughput is enhanced as $\eta$ increases. Moreover, we notice from Fig. \ref{Fig2} that increasing the capacity of the SU's battery, i.e., $E_{\max}$, enhances its throughput. These results are intuitive and match our expectations.

\begin{figure}
\begin{center}
\includegraphics[width=1\columnwidth , height=0.7\columnwidth]{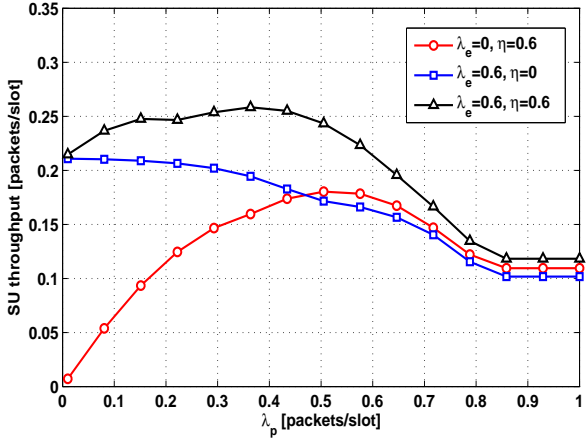}
\caption{SU throughput versus $\lambda_{\rm p}$ for different energy harvesting techniques.} \label{Fig4}
\end{center}
\vspace{-5mm}
\end{figure}

We proceed next with studying the power allocation policy employed at the PU and how it affects the throughput of both primary and secondary links. As explained in Section \ref{system_model}, the PU's transmit power depends on the quality of the link between itself and its destination, i.e., $\rm \sigma_{ppd}$. In Fig. \ref{Fig3}, we remain focusing on the case of RF energy harvesting via setting $\lambda_{\rm e}=0$. We plot the throughput of both the PU and SU versus $\rm \sigma_{ppd}$ for $\eta=0.6$, $\lambda_{\rm p}=0.4$, $E_{\max}=10$ energy packets and $P_{\mathcal{M}}=1.76$ dBm. As expected, the PU's throughput is a monotonically increasing function of $\rm \sigma_{ppd}$. From (\ref{PU_thrpt}), we note that $\mu_{\rm p}$ increases as $\rm \sigma_{ppd}$ increases. Thus, when $\mu_{\rm p}>\lambda_{\rm p}$, the PU throughput saturates at $\lambda_{\rm p}=0.4$. On the other hand, the SU throughput is enhanced initially as $\rm \sigma_{ppd}$ increases. This is attributed to the increase in the frequency of primary RF transmissions and hence, the amount of energy packets harvested by the SU increases resulting in enhancing its throughput. However, as $\rm \sigma_{ppd}$ continues to increase, the behavior of the SU throughput is reversed. This originates from the adaptive power allocation policy employed at the PU. As the quality of the link $\rm p \rightarrow pd$ is enhanced, the PU can guarantee successful packet transmissions at lower average power levels. Thus, as the PU reduces its transmission power, the amount of energy harvested by the SU is reduced yielding the reduction in its throughput.

In Fig. \ref{Fig4}, we show the resulting SU throughput for different energy harvesting scenarios at $E_{\max}=6$ energy packets. We consider each energy harvesting technique independently, i.e., natural resources and RF energy harvesting, then we combine both of them. The case of harvesting energy from natural resources only corresponds to $\eta=0$. Since the amount of energy harvested from natural resources is independent of the PU's activity, the energy packets arrivals at $Q_{\rm e}$ does not change as $\lambda_{\rm p}$ varies. Therefore, the SU throughput changes with $\lambda_{\rm p}$ only due to the variation of the availability of time slots in which the PU is inactive. Consequently, SU throughput decreases as $\lambda_{\rm p}$ increases which is evident from Fig. \ref{Fig4}. On the other hand, Fig. \ref{Fig4} shows the SU performance under RF energy harvesting only, i.e., $\lambda_{\rm e}=0$, that has already been explained in the comments on Fig. \ref{Fig2}. Finally, from Fig. \ref{Fig4}, we can notice the performance gain originating from combining both energy harvesting techniques.

\begin{figure}
\begin{center}
\includegraphics[width=1\columnwidth , height=0.7\columnwidth]{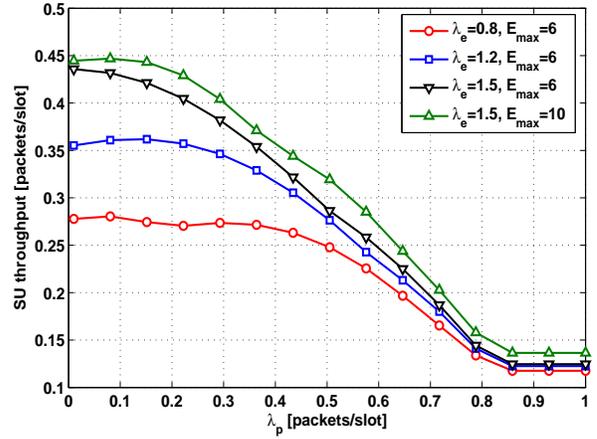}
\caption{SU throughput versus $\rm \lambda_{p}$ for different values of $\lambda_{\rm e}$ and $E_{\max}$.}\label{Fig5}
\end{center}
\vspace{-5mm}
\end{figure}

In Fig. \ref{Fig5}, we investigate the dependence of the SU throughput on energy packets' arrival rate at $Q_{\rm e}$, $\lambda_{\rm e}$, and the maximum capacity of $Q_{\rm e}$, $E_{\max}$, at $\eta=0.2$. We note that the SU throughput is enhanced as both $\lambda_{\rm e}$ in energy packets/slot and $E_{\max}$ in packets increase. This is attributed to the increase in the availability of energy at the SU required to support its packets transmission.

\section{Conclusion}\label{conclusion}
In this work, we have investigated the maximum achievable throughput for an SU sharing the spectrum with a PU. The SU has a limited-capacity battery and is equipped with energy harvesting capability. We assume that the SU harvests energy from primary RF transmissions as well as through natural resources. The obtained results reveal the promises of employing RF energy harvesting. We have showed that the SU throughput can be enhanced as the arrival rate to the primary queue increases. This is attributed to the increase in the amount of energy that the SU harvests from primary RF transmissions. Thus, we conclude that a win-win scenario can be approached in which the performance of both users is enhanced simultaneously.

\section*{Appendix A}
For simplicity, we define new variables with which we proceed in the derivation. Let $X=h_{\rm ps}$, $Y=h_{\rm ppd}$ and $Z=X/Y$. The first step towards solving (\ref{Z}) requires the computation of $F(z)=\Pr\{Z \leq z , Y \geq a\}$ for an arbitrary $z \geq 0$. After some manipulation, we have
\begin{equation}\label{17}
F(z)=\int_{a}^{\infty} \int_{0}^{zy} P_{X,Y}(x,y) dx dy
\end{equation}
where $P_{X,Y}(\cdot,\cdot)$ denotes the joint probability density function (PDF) of $X$ and $Y$. Since $X$ and $Y$ are independent random variables, their joint PDF is given by the product of their marginal distributions, i.e., $P_{X,Y}(x,y)=P_{X}(x)P_{Y}(y)$. Following the Rayleigh fading assumption, we know that $P_{X}(x)=\lambda_{x}\exp(-x \lambda_{x})$ and $P_{Y}(y)=\lambda_{y}\exp(-y \lambda_{y})$, where $\lambda_{x}$ and $\lambda_{y}$ are given by $1/\sigma_{\rm ps}$ and $1/\sigma_{\rm ppd}$, respectively. Substituting in (\ref{17}) and solving the double integral, we get
\begin{equation}
\begin{split}
F(z)\!&=\!\int_{a}^{\infty} \int_0^{zy} \lambda_x \exp(-\lambda_x x) \lambda_y \exp(-\lambda_y y) dx dy\\& \!=\! \int_a^\infty \lambda_x \lambda_y \exp(-\lambda_y y) \int_{0}^{zy}\! \exp(-\lambda_x x) dx\\& \!=\!\lambda_x \lambda_y \int_a^ \infty \! \! \exp(- \lambda_y y) \Big[ \frac{\exp(-\lambda_x x)}{-\lambda_x}\Big]_0^{zy} dy\\& \!=\!\lambda_y \int_a^ \infty\exp(- \lambda_y y) \Big[ {1-\exp(-\lambda_x zy)}\Big]_0^{zy} dy \\& \!=\!\lambda_y \int_a ^\infty\!\!\exp(-\lambda_y y)\!-\!\exp(-y [\lambda_y\!+\!\lambda_x z]) dy\\& \!=\!  \Bigg[\!-\exp(-\lambda_y y)\!+\!\frac{\lambda_y}{\lambda_y\!+\!\lambda_x z}\exp(-y [\lambda_y\!+\!\lambda_x z])\Bigg]_a^\infty\\&=
%
   \exp(-\lambda_y a)\Bigg[\!1\!-\!\frac{\lambda_y}{\lambda_y\!+\!\lambda_x z}\exp(-a \!\lambda_x z)\Bigg].
\end{split}
\end{equation}

\bibliographystyle{IEEEtran}
\bibliography{IEEEabrv,term_bib}

\begin{thebibliography}{10}
\providecommand{\url}[1]{#1}
\csname url@samestyle\endcsname
\providecommand{\newblock}{\relax}
\providecommand{\bibinfo}[2]{#2}
\providecommand{\BIBentrySTDinterwordspacing}{\spaceskip=0pt\relax}
\providecommand{\BIBentryALTinterwordstretchfactor}{4}
\providecommand{\BIBentryALTinterwordspacing}{\spaceskip=\fontdimen2\font plus
\BIBentryALTinterwordstretchfactor\fontdimen3\font minus
  \fontdimen4\font\relax}
\providecommand{\BIBforeignlanguage}[2]{{%
\expandafter\ifx\csname l@#1\endcsname\relax
\typeout{** WARNING: IEEEtran.bst: No hyphenation pattern has been}%
\typeout{** loaded for the language `#1'. Using the pattern for}%
\typeout{** the default language instead.}%
\else
\language=\csname l@#1\endcsname
\fi
#2}}
\providecommand{\BIBdecl}{\relax}
\BIBdecl

\bibitem{f1}
Z.~Hasan, H.~Boostanimehr, and V.~K. Bhargava, ``Green cellular networks: A
  survey, some research issues and challenges,'' \emph{Communications Surveys
  \& Tutorials, IEEE}, vol.~13, no.~4, pp. 524--540, 2011.

\bibitem{lu2014dynamic}
X.~Lu, P.~Wang, N.~Dusit, and H.~Ekram, ``Dynamic spectrum access in cognitive
  radio networks with rf energy harvesting,'' \emph{arXiv preprint
  arXiv:1401.3502}, 2014.

\bibitem{mikeka2011design}
C.~Mikeka and H.~Arai, ``Design issues in radio frequency energy harvesting
  system,'' \emph{Sustainable Energy Harvesting Technologies--Past, Present and
  Future}, pp. 235--256, 2011.

\bibitem{f3}
A.~Cuadras, M.~Gasulla, and V.~Ferrari, ``Thermal energy harvesting through
  pyroelectricity,'' \emph{Sensors and Actuators A: Physical}, vol. 158, no.~1,
  pp. 132--139, 2010.

\bibitem{hoang2009opportunistic}
A.~Hoang, Y.~Liang, D.~Wong, Y.~Zeng, and R.~Zhang, ``Opportunistic spectrum
  access for energy-constrained cognitive radios,'' \emph{IEEE Trans. Wireless
  Commun.}, vol.~8, no.~3, pp. 1206--1211, Mar. 2009.

\bibitem{sharma2010optimal}
V.~Sharma, U.~Mukherji, V.~Joseph, and S.~Gupta, ``Optimal energy management
  policies for energy harvesting sensor nodes,'' \emph{IEEE Trans. Wireless
  Commun.}, vol.~9, no.~4, pp. 1326--1336, Apr. 2010.

\bibitem{ho2010optimal}
C.~Ho and R.~Zhang, ``Optimal energy allocation for wireless communications
  powered by energy harvesters,'' in \emph{Proc. IEEE ISIT}, Jun. 2010, pp.
  2368--2372.

\bibitem{yang2010transmission}
J.~Yang and S.~Ulukus, ``Transmission completion time minimization in an energy
  harvesting system,'' in \emph{Proc. IEEE CISS}, Mar., pp. 1--6.

\bibitem{yang2010optimal}
------, ``Optimal packet scheduling in an energy harvesting communication
  system,'' \emph{IEEE Trans. Commun.}, vol.~60, no.~1, pp. 220--230, Jan.
  2012.

\bibitem{tutuncuoglu2010optimum}
K.~Tutuncuoglu and A.~Yener, ``Optimum transmission policies for battery
  limited energy harvesting nodes,'' \emph{IEEE Trans. Wireless Commun.},
  vol.~11, no.~3, pp. 1180--1189, Mar. 2012.

\bibitem{survey}
S.~Sudevalayam and P.~Kulkarni, ``Energy harvesting sensor nodes: Survey and
  implications,'' \emph{IEEE Communications Surveys and Tutorials}, vol.~13,
  no.~3, pp. 443--461, 2011.

\bibitem{pappas}
N.~Pappas, J.~Jeon, A.~Ephremides, and A.~Traganitis, ``Optimal utilization of
  a cognitive shared channel with a rechargeable primary source node,'' in
  \emph{JCN}, vol.~14, no.~2, Apr. 2012, pp. 162--168.

\bibitem{wimob}
A.~{El Shafie} and A.~Sultan, ``Optimal random access and random spectrum
  sensing for an energy harvesting cognitive radio,'' in \emph{Proc. IEEE
  WiMob}, Oct. 2012, pp. 403--410.

\bibitem{ourletter}
------, ``Optimal random access for a cognitive radio terminal with energy
  harvesting capability,'' \emph{IEEE Commun. Lett.}, vol.~17, no.~6, pp.
  1128--1131, 2013.

\bibitem{sultan}
A.~Sultan, ``Sensing and transmit energy optimization for an energy harvesting
  cognitive radio,'' \emph{IEEE Wirel. Commun. Lett.}, vol.~1, no.~5, pp.
  500--503, Oct. 2012.

\bibitem{wcmpaper}
A.~{El Shafie}, T.~Khattab, A.~El-Keyi, and M.~Nafie, ``On the coexistence of a
  primary user with an energy harvesting secondary user: A case of cognitive
  cooperation,'' {To} appear in {\it Wireless Communications and Mobile
  Computing}, 2014.

\bibitem{gc2013}
A.~{El Shafie} and A.~Sultan, ``Optimal selection of spectrum sensing duration
  for an energy harvesting cognitive radio,'' in \emph{Proc. IEEE GLOBECOM},
  2013, pp. 1020--1025.

\bibitem{krikidis2012stability}
I.~Krikidis, T.~Charalambous, and J.~Thompson, ``Stability analysis and power
  optimization for energy harvesting cooperative networks,'' \emph{IEEE Signal
  Process. Lett.}, vol.~19, no.~1, pp. 20--23, Jan. 2012.

\bibitem{le2008efficient}
T.~Le, K.~Mayaram, and T.~Fiez, ``Efficient far-field radio frequency energy
  harvesting for passively powered sensor networks,'' \emph{IEEE J. of
  Solid-State Circuits}, vol.~43, no.~5, pp. 1287--1302, 2008.

\bibitem{f7}
S.~Park, H.~Kim, and D.~Hong, ``Cognitive radio networks with energy
  harvesting,'' 2013.

\bibitem{f8}
S.~Lee, R.~Zhang, and K.~Huang, ``Opportunistic wireless energy harvesting in
  cognitive radio networks,'' \emph{IEEE Trans. Wireless Commun.}, vol.~12,
  no.~9, pp. 4788--4799, September 2013.

\bibitem{f9}
S.~Park, J.~Heo, B.~Kim, W.~Chung, H.~Wang, and D.~Hong, ``Optimal mode
  selection for cognitive radio sensor networks with rf energy harvesting,'' in
  \emph{IEEE PIMRC}, Sept 2012, pp. 2155--2159.

\bibitem{krikidis2009protocol}
I.~Krikidis, J.~Laneman, J.~Thompson, and S.~McLaughlin, ``Protocol design and
  throughput analysis for multi-user cognitive cooperative systems,''
  \emph{IEEE Trans. Wireless Commun.}, vol.~8, no.~9, pp. 4740--4751, 2009.

\bibitem{krikidis2010stability}
I.~Krikidis, N.~Devroye, and J.~Thompson, ``Stability analysis for cognitive
  radio with multi-access primary transmission,'' \emph{IEEE Trans. Wireless
  Commun.}, vol.~9, no.~1, pp. 72--77, 2010.

\end{thebibliography}
\end{document}